\documentstyle[12pt]{article}
\renewcommand{\theequation}{\arabic{section}.\arabic{equation}}
\newcounter{saveeqn}

\begin{document}

\begin{titlepage}\hfill HD-THEP-99-32\\[5 ex]

\begin{center}{\Large\bf Analytic approach to confinement \\ and
monopoles in lattice {$\bf SU(2)$}} \\[5 ex]

{\bf Dieter Gromes}\\[3 ex]Institut f\"ur
Theoretische Physik der Universit\"at Heidelberg\\ Philosophenweg 16,
D-69120 Heidelberg \\ E - mail: d.gromes@thphys.uni-heidelberg.de \\
 \end{center} \vspace{2cm}

{\bf Abstract: } We extend the approach of Banks, Myerson, and Kogut
for the calculation of the Wilson loop in lattice $U(1)$ to the
non-abelian $SU(2)$ group. The original degrees of freedom of the
theory are integrated out, new degrees of freedom are introduced in
several steps. The centre group $Z_2$ enters automatically through the
appearance of a field strength tensor $f_{\mu \nu }$, which takes on
the values 0 or 1 only. It obeys a linear field equation with the loop
current as source. This equation implies that $f_{\mu \nu }$ is non
vanishing on a two-dimensional surface bounded by the loop, and
possibly on closed surfaces. The two-dimensional surfaces have a
natural interpretation as strings moving in euclidean time.
In four dimensions we recover the dual Abrikosov string of
a type II superconductor, i.e. an electric string encircled by a
magnetic current. In contrast to other types of monopoles found in the
literature, the monopoles and the associated magnetic currents are
present in every configuration. With some plausible, though not
generally conclusive, arguments we are directly led to the area law
for large loops.

 \vfill \centerline{August 1999, revised October 1999}

\end{titlepage}

\section{Introduction}

It is now widely accepted, that confinement is due to the formation of
a color electric string, and that magnetic monopoles play an essential
role in this context. Up to now there is a lively activity in this
field, illuminating the phenomenon from various sides. A particularly
appealing approach is the one by Banks, Myerson, and Kogut \cite{bmk}
which is now more than 20 years old. They considered the partition
function for an electric current loop and derived step by step the
appearance of monopoles by integrating out the original degrees of
freedom and introducing new ones. The possibility to do this was
restricted to the lattice U(1) model with the Villain action
\cite{vil} and some other simple models. The authors also remark, that
the techniques do not generalize simply to non-abelian theories.

We will start by applying and generalizing the methods of ref
\cite{bmk}. In a first step the $SU(2)$ matrices on the links are
explicitly parametrized by three angles $\psi ,\vartheta ,\varphi $.
An appropriate decomposition of the $SU(2)$ matrices allows the
calculation of the trace in the plaquette action. After an expansion
of exponentials into modified Bessel functions the integrations over
the link angles can be performed. They lead to constraints for the new
variables which were introduced in the expansions. Most of these
variables are irrelevant and the summations can be performed after a
suitable transformation.

After this has been done, we are left with several integer variables
which are restricted by constraint equations. The most important one
is a $Z_2$ field strength tensor $f_{\mu \nu }$. It lives on
plaquettes and is either 0 or 1. This tensor obeys an inhomogeneous
linear field equation with the loop current as source. The solutions
of the equation have a simple geometrical interpretation. The tensor
$f_{\mu \nu }$ is non vanishing on a two-dimensional surface bounded
by the loop, and possibly on closed two-dimensional surfaces. These
surfaces have a natural interpretation as strings moving in euclidean
time. There is a string which connects the two charges associated with
the loop, and possibly a number of additional closed strings. The
situation is particularly transparent for planar loops, where the
layer on the minimal surface, corresponding to the straight string,
plays a special role. For large loops we can use some general
arguments and reasonings from statistical mechanics, like additivity
of the free energy, and obtain the area law. The subtle question,
whether a finite string tension survives in the thermodynamic limit
and in the continuum limit, needs additional investigations.

Our approach is purely analytical and non-perturbative, no gauge
fixing was performed, and $\beta $ was kept arbitrary.
Lengthy calculations were done with the help of Mathematica. Nowhere
any physical picture of what we expect was put in. It is the formalism
itself which automatically leads to the appearance of a plaquette
variable $f_{\mu \nu }$ which is naturally associated with the world
sheet of a string. If one could perform the integrations and
summations over the remaining parameters one would obtain the explicit
string action. Even without doing this, the formalism clearly shows
the appearance of the string picture and the origin of the confinement
mechanism.

\setcounter{equation}{0}\addtocounter{saveeqn}{1}%

\section{The partition function }

We are interested in the expectation value of a Wilson loop $W$,
characterized by a closed current loop $J$:

\begin{equation} Z[J]=\int Tr[W(J)]\exp[\frac{\beta }{2}\sum_{p_{\rho
\nu }}TrU_{\rho \nu }(p)] {\cal D}[U].\end{equation}
The sum runs over all plaquettes $p_{\rho \nu }$ (with $\rho <\nu $),
while $U_{\rho \nu } (p)$ is the product of the four $SU(2)$-matrices
on the links of the plaquette. We will perform rather extensive
manipulations in the following, therefore we fix our notation here:

\begin{eqnarray} & & p,q,r \mbox{\quad denote lattice points},
\nonumber\\
& & \mu ,\nu ,\rho ,\lambda =1,\cdots ,d \mbox{\quad denote
space directions,} \nonumber\\
& & p\pm\mu \mbox{\quad is the lattice point next to }p
\mbox{\quad in positive or negative }\mu\mbox{-direction,}
\nonumber\\
& & p_\mu \mbox{\quad denotes the link connecting $p$ with $p+\mu $},
\\
& & p_{\mu \nu } \mbox{\quad with \quad $\mu <\nu$
\quad is the plaquette determined by the links $p_\mu $ and
$p_\nu $,}\nonumber\\
& & \Theta ^a_\mu (p) =(\psi ,\vartheta ,\varphi
)_\mu (p)
 \mbox{\quad denotes three angles which parametrize the SU(2)
  matrices} \nonumber\\
& & \mbox{on the link $p_\mu $, indices $a,b$ generally
run over $\psi,\vartheta,\varphi $},\nonumber\\
& & U_\mu (p) \equiv U(\Theta_\mu ^a(p)) \mbox{ is the $SU(2)$ matrix
on the link $p_\mu $},\nonumber\\
& & U_{\rho \nu }(p) =U_\rho (p)U_\nu (p+\rho ) U_\rho ^+(p+\nu
)U_\nu ^+(p) \mbox{ is the plaquette variable}.\nonumber
\end{eqnarray}
As a first step we have to choose a parametrization for
the link variables $U_\mu (p)$ in order to be able to do the group
integrations. We proceed similarly as in previous work
which applied the optimized $\delta
$-expansion on the lattice \cite{bj}, \cite{bg}. In our case the
Euler parametrization is the most appropriate, i.e. we use

\begin{equation} U = e^{i\sigma _3\psi } e^{i\sigma _2\vartheta }
e^{i\sigma _3\varphi }. \end{equation}
The Haar measure is proportional to $\sin(2\vartheta)$.
A possible choice of the parameter space is $-\pi <\psi <\pi
,\quad 0 <\varphi <\pi ,\quad 0<\vartheta<\pi /2$.

For the following it is convenient to extend this region. All
integrals which appear contain functions of the $U_{\rho \nu
}(p)$ which are periodic in the Euler angels. So we may use some
symmetry relations which are easily seen from the decomposition of $U$
into 1 and the $\sigma _m$. The shift $\varphi \rightarrow \varphi
-\pi ,\psi \rightarrow \psi -\pi $ leaves $U$ invariant. Therefore one
can extend the $\varphi $-integration into the interval from $-\pi $
to $\pi $, thus covering the group manifold twice. The further
symmetry $\vartheta \rightarrow -\vartheta , \psi \rightarrow \psi
-\pi /2, \varphi \rightarrow \varphi +\pi /2$ allows to extend the
$\vartheta $-integration to the interval $-\pi /2<\vartheta <\pi /2$
if we continue the Haar measure as even function. Finally the symmetry
$\vartheta \rightarrow \vartheta -\pi ,\psi \rightarrow \psi -\pi $
allows the extension of the $\vartheta $-integration to the full
interval. Therefore we take

\begin{equation} H(\vartheta) = \frac{\pi }{2}|\sin(2\vartheta)|
\end{equation}
as Haar measure in the following, and use the common boundaries $-\pi
<\psi , \vartheta ,\varphi <\pi $.

In fig. 1 we show the notation for the link variables of the plaquette
$p_{\rho \nu }$.

\unitlength1ex
\begin{picture}(80,40)
\put(27,5){\vector(1,0){26}}
\put(55,7){\vector(0,1){26}}
\put(27,35){\vector(1,0){26}}
\put(25,7){\vector(0,1){26}}
\put(36,2){$U_\rho (p)$}
\put(56,20){$U_\nu (p+\rho )$}
\put(34,37){$U^+_\rho (p+\nu )$}
\put(17,20){$U^+_\nu (p)$}
\put(56,5){$\rho $}\put(23,37){$\nu $}\put(20,3){$(p)$}
\end{picture}

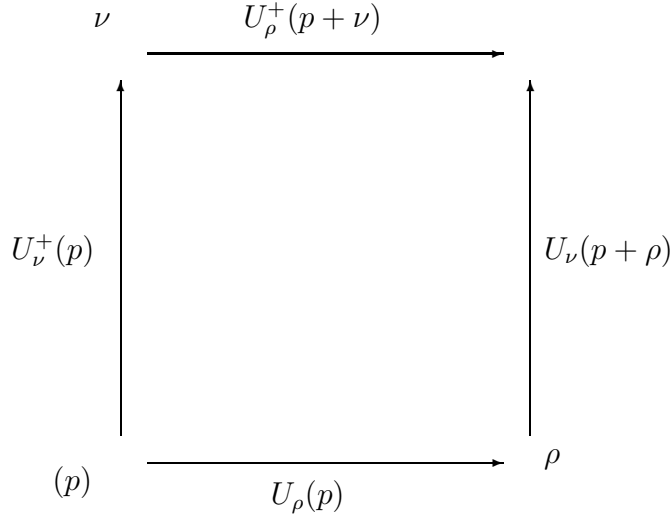
{\bf Fig.1.} The plaquette $p_{\rho \nu }$ and the link variables.

\quad

The parametrization according to (2.3) becomes

\begin{eqnarray} U_\rho (p) & = & e^{i\sigma _3\psi _\rho (p)}\;
e^{i\sigma _2\vartheta _\rho (p)} \;e^{i\sigma _3\varphi _\rho (p)},
\nonumber\\
U_\nu (p+\rho ) & = & e^{i\sigma _3\psi _\nu (p+\rho )}\;
e^{i\sigma _2\vartheta _\nu (p+\rho )} \;e^{i\sigma _3\varphi _\nu
(p+\rho )},\nonumber\\
U^+_\rho (p+\nu ) & = & e^{-i\sigma _3\varphi _\rho (p+\nu )}
\;e^{-i\sigma _2\vartheta _\rho (p+\nu )} \;e^{-i\sigma _3\psi
_\rho (p+\nu )},\\
U^+_\nu (p) & = & e^{-i\sigma _3\varphi  _\nu (p)} \;e^{-i\sigma
_2\vartheta _\nu (p)} \;e^{-i\sigma _3\psi _\nu (p)}. \nonumber
\end{eqnarray}
An appropriate technique for the further procedure, which was also
extensively used in \cite{bj} and \cite{bg}, is the splitting of
the matrix exponentials into sums of ordinary exponentials times
projection operators, in general

\begin{equation} e^{\pm i\sigma _m\alpha } = \sum_{s=\pm 1} e^{\pm i s
\alpha } P_s(m),\mbox{\quad with\quad }P_s(m) = \frac{1}{2 }(1+s\sigma
_m).\end{equation}
The link variables $U$ are parametrized as in (2.5), the plaquette
variable $U_{\rho \nu }(p)$ therefore contains 12 factors. To every
factor we apply the decomposition (2.6). This means that we need 12
summation indices $s[\Theta ] =\pm 1$, associated with the twelve
angles in (2.5). At the end $TrU_{\rho \nu }(p)$ becomes a product of
12 factors, each being a sum of two terms of the type (2.6). So in
total we have a sum of $2^{12}$ terms. Each term of the sum consists
of a trace $T$ of a product of 12 projectors $P_s(m)$, multiplied by
an exponential.

At the four corners of the plaquette one has a product of two
projectors $P_s(3)P_{s'}(3)$. They both project on the same subspace,
therefore we get a non-vanishing result only if the neighboring
parameters $s$ and $s'$ agree. So we must have $s[\varphi
_\rho (p)] =s[\psi _\nu (p+\rho )],$ $s[\varphi _\nu (p+\rho )]
=s[\varphi _\rho (p+\nu )], s[\psi _\rho (p+\nu )] =s[\varphi _\nu
(p)], s[\psi _\nu (p)] =s[\psi _\rho (p)]$, there are in fact
not 12, but only 8 independent parameters.

We enumerate the remaining 8 independent
parameters $s[\Theta]$ in a consecutive way, starting with
$s[\vartheta _\rho (p)]
$. In (2.7) we give the parameters $s_i$, together with the angles to
which they refer. According to the remarks above, the $s_i$ with
even $i$ belong to two angles.

\begin{eqnarray}
s_1 & \Leftrightarrow & \vartheta _\rho (p), \quad \quad \; \; \:
s_2  \Leftrightarrow  \varphi _\rho (p),\psi _\nu (p+\rho ),
\nonumber\\
s_3 & \Leftrightarrow & \vartheta _\nu (p+\rho ), \quad
s_4  \Leftrightarrow  \varphi _\nu (p+\rho ),\varphi _\rho (p+\nu ),
\nonumber\\
s_5 & \Leftrightarrow & \vartheta _\rho (p+\nu ), \quad
s_6  \Leftrightarrow  \psi  _\rho (p+\nu ),\varphi  _\nu (p), \\
s_7 & \Leftrightarrow & \vartheta _\nu (p), \quad \quad \; \; \:
s_8  \Leftrightarrow  \psi  _\nu (p),\psi _\rho (p). \nonumber
\end{eqnarray}
The number of terms in the expansion of $TrU_{\rho \nu }(p)$ has
now been reduced from $2^{12}$ to $2^8$. We enumerate these terms by
an index $n$, the order in which this is done needs not to be
specified yet. Thus each $s_i$ becomes a function of $n$, we
denote it by $s_{in}$. The traces also depend upon
$n$, we denote them by $T_n$. Finally one can write
$TrU_{\rho \nu }(p)$ in the following form:

\begin{equation} TrU_{\rho \nu }(p) =Tr[U_\rho (p)U_\nu (p+\rho )
U_\rho ^+(p+\nu )U_\nu ^+(p)] = \sum_{n=1}^{2^8} T_n\exp[iA_{\rho
\nu }^n(p;\Theta ,s)]. \end{equation}
Here

\begin{eqnarray}
A_{\rho \nu }^n(p;\Theta ,s)  & = &
s_{8n} \psi _\rho (p) + s_{1n} \vartheta _\rho (p) + s_{2n} \varphi _
\rho (p) \nonumber\\
& & + s_{2n} \psi _\nu (p+\rho ) + s_{3n} \vartheta _\nu (p+\rho )
+ s_{4n} \varphi _\nu (p+\rho ) \\
& & - s_{4n} \varphi _\rho (p+\nu ) - s_{5n} \vartheta _\rho (p+\nu )
- s_{6n} \psi _\rho (p+\nu ) \nonumber\\
& & - s_{6n} \varphi _\nu (p) - s_{7n} \vartheta _\nu (p)
- s_{8n} \psi _\nu(p). \nonumber
\end{eqnarray}
The computation of the non-vanishing $2^{8}$ traces $T_n$ shows that
they all have the values $\pm 1/16$. This structure is easily
understood. Applying the projectors $P_s(3)$ and $P_s(2)$ to a given
vector successively leads to alternating projections on different
subspaces. Each projection gives a factor $\pm 1/\sqrt{2}$, with the
sign depending upon the $s_{in}$. There are 144 values of $n$ with
$T_n=1/16$ and 112 with $T_n=-1/16$. Obviously $\sum_nT_n=2$ as it
should. A further simplification is obtained from the symmetry
$T_n(s_{in}) = T_n(-s_{in})$, which is easily seen by inserting
$\sigma _1\sigma _1$ between all projectors and using $\sigma
_1P_s(m)\sigma _1 = P_{-s}(m)$ for $m=2,3$. Together with the symmetry
$A_{\rho \nu }^n(p;\Theta ,s_{in}) = - A_{\rho \nu }^n(p;\Theta
,-s_{in})$ and the reality of the trace in (2.8) we can therefore fix,
e.g. $s_{8n} =1$, multiply (2.8) by 2 on the rhs, and restrict the sum
over $n$ to the $2^7$ values $s_{1n} ,\cdots ,s_{7n}$. It is
convenient to introduce sign factors $\epsilon _n=\pm 1$ and write

\begin{equation} T_n = \frac{\epsilon _n}{16}  =
Tr[(P_{s_{1n}} (2) P_{s_{2n}} (3)) \;
(P_{s_{3n}}(2)P_{s_{4n}}(3)) \;
(P_{s_{5n}}(2)P_{s_{6n}}(3)) \;
(P_{s_{7n}}(2) P_{s_{8n}=1}(3))].
\end{equation}
According to the previous remarks we dropped redundant projectors and
fixed $s_{8n}=1$. We may now write (2.8) as sum over the restricted
set $n$ in the form

\begin{equation} Tr U_{\rho \nu }(p) = \frac{1}{8}\sum_{n=1}^{2^7}
 \epsilon _n \cos[A_{\rho \nu }^n(p;\Theta ,s)]. \end{equation}
In order to perform the integrations over all the angles
$\psi ,\vartheta,\varphi $ we proceed essentially as in \cite{bmk}.
The various exponentials are expanded into a series of modified Bessel
functions according to the formula

\begin{equation} \exp[\epsilon z \cos A] = \sum_{l=-\infty}^{\infty}
\epsilon ^l I_l(z) e^{ilA} \mbox{\quad for }\epsilon =\pm 1.
\end{equation}
The replacement $\epsilon \rightarrow
-\epsilon $ is obviously equivalent to the shift $A\rightarrow A+\pi
$. Using (2.11) and (2.12) we thus find

\begin{eqnarray} \exp[\frac{\beta }{2}\sum_{p_{\rho \nu }}
TrU_{\rho \nu }(p)]
& = & \exp\left[ \frac{\beta }{16} \sum_{p_{\rho \nu} ,n}\epsilon _n
\cos [A_{\rho \nu }^n(p;\Theta ,s)]\right] \nonumber\\
& = & \prod_{p_{\rho \nu },n}\exp\left[ \frac{\beta }{16}
\epsilon _n \cos [A_{\rho \nu }^n(p;\Theta ,s)]\right] \nonumber\\
& = & \prod_{p_{\rho \nu },n}
\sum_{l=-\infty}^\infty (\epsilon _n)^l I_l(\frac{\beta }{16}) \exp
[il A_{\rho \nu }^n(p;\Theta ,s)]\\
& = & \sum_{l_{\rho \nu }^n(p)} \prod_{p_{\rho \nu },n} (\epsilon _n)
^{l_{\rho \nu }^n(p)} I_{l_{\rho \nu
}^n(p)}(\frac{\beta }{16})\exp[il_{\rho\nu }^n(p)
A_{\rho \nu} ^n(p;\Theta ,s)]\nonumber. \end{eqnarray}
In the last step we exchanged the order of the product and the sum.
The summation parameters $l$ must then be distinguished by indices
referring to the corresponding plaquette $p_{\rho \nu }$ and the
index $n$; the $l_{\rho \nu }^n(p)$ run independently from
$-\infty$ to $\infty$.

We can now write down the expectation value for a Wilson loop. We
characterize it by a closed current loop $J_\lambda (q)$ which is $\pm
1$ if the current runs in or against the direction of the link
$q_\lambda $, and 0 otherwise. For simplicity we exclude loops which
run multiply through some links. For the
calculation of the trace $Tr[W(J)]$ we use again the decomposition
(2.6) for the link variables on the loop, parametrized by (2.3). For
every link $q_\lambda $ on the loop we have two parameters
$\hat{s}^b_\lambda (q) = \pm 1$ (not three, because again neighboring
projectors have to coincide), the whole set of these will be denoted
by $\hat{s}$. For a loop of lengt $L$ we have $2L$ parameters
$\hat{s}^b_\lambda (q) = \pm 1$ and in total a sum of $2^{2L}$ terms.
We count these by an index $\hat{n}$, and denote the parameters by
$s_{\lambda \hat{n}}^b(q)$. The traces are called
$W_{\hat{n}}$. The $W_{\hat{n}}$ have a similar structure as the
$T_n$; their values are $\pm1/2^L$, and $\sum _{\hat{n}}W_{\hat{n}}
=2$. The partition function now reads

\begin{eqnarray}
Z[J] & = & \int \sum_{\hat{n}} W_{\hat{n}} \exp[i\sum_{q\lambda
b}J_\lambda (q) \hat{s}_{\lambda \hat{n}}^b(q) \Theta _\lambda
^b(q)]\nonumber\\
& & \sum_{l_ {\rho \nu }^n(p)} \left(\prod_{p_{\rho \nu },n}
(\epsilon _n)^{l_{\rho \nu }^n(p)} I_{l_{\rho \nu }^n(p)}
(\frac{\beta }{16})\exp[i l_{\rho \nu }^n(p)A_{\rho \nu
}^n(p;\Theta ,s)] \right)\nonumber\\
& & \prod_{r\mu }\left(H(\vartheta _\mu (r)) \prod_a \frac{d
\Theta _\mu ^a(r)}{2\pi }\right).\end{eqnarray}
Obviously the first sum over
$\hat{n}$ is the expansion of the loop. The sum over $l_{\rho \nu
}^n(p)$ in the second line contains the action transformed as in
(2.13), while the product over $r,\mu $ in the third line contains the
integrations together with the Haar measure. We may factorize the
product over $p_{\rho \nu },n$ and apply the addition theorem for the
exponentials. This results in

\begin{eqnarray} Z[J] & = & \int \sum_{\hat{n}} W_{\hat{n}}
\exp[i\sum_{q\lambda  b}J_\lambda  (q)\hat{s}_{\lambda \hat{n}}
^b(q)\Theta_\lambda ^b(q)]\nonumber\\
& & \sum_{l_{\rho \nu }^n(p)} \left(\prod_{p_{\rho \nu },n}
(\epsilon _n)^{l_{\rho \nu }^n(p)} I_{l_{\rho \nu }^n(p)}
(\frac{\beta }{16}) \right) \exp [i \sum_{p,\rho<\nu ,n}
l_{\rho \nu }^n(p)A_{\rho \nu}^n(p;\Theta ,s)]\nonumber\\
& & \prod_{r\mu }\left( H(\vartheta _\mu (r))\prod_a \frac{d\Theta_\mu
^a(r)}{2\pi }\right). \end{eqnarray}
We are now ready to perform the angular integrations over $\Theta_\mu
^a(r)$, remembering the definition of $A_{\rho \nu }^n(p;\Theta ,s)$
in (2.9). The integrations over $\psi _\mu (r)$ and $\varphi _\mu (r)$
lead to a Kronecker-$\delta $ which gives a constraint, while the
$\vartheta _\mu (r)$ integration involves the Haar measure (2.4) and
leads to a more complicated function. We shall also call it a
constraint for simplicity. It is convenient to define a symbol $\delta
^a(C)$ for integer $C$ by

\begin{equation} \delta ^{a}(C) = \left\{ \begin{array}{l} \int _{-\pi
}^\pi e^{iC\psi } \frac{d\psi }{2\pi } = \delta
_{C,0} \mbox{ for } a=\psi ,\varphi ,\\
\int _{-\pi }^\pi H(\vartheta )e^{iC\vartheta } \frac{d\vartheta
}{2\pi } = \left\{ \begin{array}{l}1/(1-C^2/4) \mbox{\quad if $C$ is a
multiple of 4},\\ 0 \mbox{\quad otherwise} \end{array}
\right\} \mbox{ for } a=\vartheta .\end{array}\right.
\end{equation}
The argument of the constraint which arises from the $\Theta _\mu
^a(r)$-integration becomes

\begin{eqnarray} C_\mu ^a(r) & \equiv &
\sum_{\nu >\mu ,n} [s_{(8,1,2)n} l_{\mu \nu }^n(r)-
s_{(6,5,4)n} l_{\mu \nu }^n(r-\nu )]\\
& - & \sum_{\nu <\mu ,n}[s_{(8,7,6)n} l_{\nu \mu }^n(r)
- s_{(2,3,4)n} l_{\nu \mu }^n(r-\nu )] + J_\mu
(r)\hat{s}_{\mu \hat{n}}^a(r), \nonumber \end{eqnarray}
where one has to use the first, second, or third subscript on $s$ for
$a=\psi ,\vartheta ,\varphi $ respectively.
So we end up with the following expression for the expectation value
of the loop:

\begin{equation} Z[J]  =
\sum_{\hat{n}} W_{\hat{n}} \sum_{l_{\rho \nu }^n(p)}
\left(\prod_{p_{\rho \nu },n} (\epsilon _n)^{l_{\rho \nu
}^n(p)} I_{l_{\rho \nu }^n(p)}(\frac{\beta }{16}) \right) \prod_{r\mu
a}  \delta ^a [C_\mu ^a(r)].
\end{equation}

\setcounter{equation}{0}\addtocounter{saveeqn}{1}%

\section{Integrating out unnecessary variables }

The constraint equations (2.17), as they stand, have a quite different
character than the corresponding ones in the abelian case in
\cite{bmk}. There one had one parameter $l_{\mu \nu }(r)$ for every
plaquette $r_{\mu \nu }$, in our case we have $2^7$ parameters
$l^n_{\mu \nu }(r)$ which are characterized by the additional
index $n$. Any attempt of a physical interpretation of the $l_{\mu
\nu }^n(r)$ would be premature at this stage.

Let us consider a fixed plaquette $r_{\mu \nu }$ for the moment, and
suppress the indices $r_{\mu \nu }$. We first look for a suitable
linear transformation from the parameters $l^n$ to new parameters
$m^i$. There are 8 combinations of the $l^n$ which play a special
role, namely the 8 sums $\sum_{n=1}^{128}s_{in}l^n$ which appear in
the constraints (2.17) (remember that $s_{8n}$ was fixed to 1). We
will choose these 8 combinations as new variables $m^i,\:i=1,\cdots
,8$, eliminate the first eight $l^n$, and keep the rest of the
parameters as they are. The $8 \times 8$ matrix made up of the
$s_{ij}$ with $i,j = 1,\cdots ,8$ will be denoted by $S$.

One may consider the $s_{in}$ as a set of $2^7$
vectors ${\bf S}_n=(s_{1n},\cdots ,s_{7n},s_{8n}=1)$ with
components $\pm 1$. Up to now we did not specify the order in which
these vectors ${\bf S}_n$ should be enumerated. It is now convenient
to choose the ${\bf S}_n$ associated with the first eight values of
$n$ in such a way that $\epsilon _n=1$ for $n=1,\cdots ,8$, with
$\epsilon _n$ the signs of the trace (2.10). This can be done in many
ways. To be definite we give our choice in the appendix. Our criteria
were a small but non-vanishing determinant of $S$ (it is 128 for the
$S$ in A.1), and a structure as transparent as possible in some of the
equations below. The transformation finally becomes

\begin{eqnarray} m^i & = & \sum_{n=1}^{128}s_{in}l^n =
\sum_{j=1}^8s_{ij}l^j + \sum_{\alpha =9}^{128}s_{i\alpha}
l^\alpha \mbox{\quad for \quad } i=1,\cdots,8,\nonumber\\
m^\alpha & = & l^\alpha \mbox{\quad for \quad } \alpha = 9,\cdots ,
128.\end{eqnarray}
In matrix form the transformation reads

\begin{equation} \left(\begin{array}{cc}m_c\\m_f\end{array}\right)
=\left(\begin{array}{cc} S & T\\0 & 1 \end{array}\right)
\left(\begin{array}{cc}l_c\\l_f\end{array}\right) ,\quad
\left(\begin{array}{cc}l_c\\l_f\end{array}\right)
=\left(\begin{array}{cc} S^{-1} &
-S^{-1}T\\0 & 1 \end{array}\right)
\left(\begin{array}{cc}m_c\\m_f\end{array}\right). \end{equation}
We have split $m$ into an eight dimensional ``constrained'' vector
$m_c$, and a 120-dimensional ``free'' vector $m_f$.
Here $S$ is the $8\times 8$ matrix defined before, with $S_{ij} =
s_{ij}$. $T$ is an $8\times120$ matrix, if we enumerate the columns
from 9 to 128 we have $T_{i\alpha } = s_{i\alpha }$.

There is an important restriction which has to be imposed on the
transformation. It is this restriction which will later on lead to the
area law. If the $l^n$ run over all integers, the same will be true
for the $m^\alpha $ with $\alpha =9,\cdots ,128$, as seen from (3.1).
But it is not true for the $m^i$ with $i=1,\cdots ,8$. From the
inversion of the transformation in (3.2) one finds that only those
$m^i$ appear, which fulfill the condition

\begin{equation} l^j = \sum_{i=1}^{8} (S^{-1})_{ji} m^i -
\sum_{i=1}^{8} (S^{-1})_{ji} \sum_{\alpha =9}^{128} s_{i\alpha }
m^\alpha \stackrel{!}{=} \mbox{ integer for }j=1,\cdots,8.
\end{equation}
A computation of the matrix elements $(S^{-1})_{ji}$ shows that they
are integer or half integer, and that the sums over the
elements of any row are integer. Because the $s_{i\alpha }$ are $\pm
1$, this implies that the second sum in (3.3) is
automatically integer. Therefore the condition simplifies to
$\sum_{i=1}^{8}(S^{-1})_{ji} m^i \stackrel{!}{=}$ integer for
$j=1,\cdots ,8$. This restriction finally becomes equivalent to

\begin{equation} m^i = \mbox{ even for all }i, \mbox{ or } m^i =
\mbox{ odd for all }i = 1,\cdots ,8. \end{equation}
Introducing the transformation (3.2) into the product in (2.18)
one obtains

\begin{equation} \sum_{l^n} \prod_n (\epsilon _n)^{l^n}
I_{l^n}(\frac{\beta }{16})
= \sum_{m^i} \sum_{m^\alpha } \left( \prod_{j=1}^8
I_{l^j(m^i,m^\alpha )} (\frac{\beta }{16}) \right) \prod_{\alpha
=9}^{128} (\epsilon _\alpha )^{m^\alpha }I_{m^\alpha }(\frac{\beta
}{16}), \end{equation}
with $l^j(m^i,m^\alpha ) = \sum_i(S^{-1})_{ji}m^i - \sum_{\alpha }
(S^{-1}T)_{j\alpha }m^{\alpha }$. The sum over $m^i$ only runs over
the subset which fulfills (3.4). The $m^\alpha $ for $\alpha
=9,\cdots,128$ do not show up in the constraints, therefore the
summations can be performed. To do this we introduce the integral
representation for all the modified Bessel functions, thereby
partially reversing the previous step of expanding the exponents.

\begin{equation} \epsilon ^l I_l(\frac{\beta }{16}) =
\int _{-\pi }^\pi \frac{d\gamma }{2\pi }
\exp[\epsilon \frac{\beta }{16} \cos \gamma + il\gamma ]
\mbox{\quad for } \epsilon =\pm 1. \end{equation}
The $m^\alpha $ appear in the exponent now, and the
summations can be performed with the help of the Poisson sum formula

\begin{equation} \sum _{m^\alpha }\exp \left[ i[\gamma _\alpha
-\sum_{j=1}^8\gamma _j(S^{-1}T)_{j\alpha }]m^\alpha \right] = 2\pi
\sum_{k_\alpha } \delta [\gamma _\alpha -\sum_{j=1}^8\gamma
_j(S^{-1}T)_{j\alpha } - 2\pi k_\alpha ]. \end{equation}
All $\gamma _\alpha $ are integrated over the interval $-\pi <\gamma
_\alpha \le\pi $, thus exactly one $k_\alpha $ contributes in the sum
on the rhs. Furthermore the $\gamma _\alpha $ appear only as
arguments of periodic cosines in the remaining part of the integrand.
Therefore we may simply use the $\delta $-functions to
eliminate the $\gamma _\alpha $ by putting $\gamma _\alpha
=\sum_{j=1}^8 \gamma _j(S^{-1}T)_{j\alpha }$ for $\alpha =9,\cdots
,128$. We thus are left with

\begin{eqnarray} \sum_{l^n} \prod_n (\epsilon _n)^{l^n}
I_{l^n}(\frac{\beta }{16})
= \sum_{m^i} \int \frac{d\gamma _1}{2\pi }\cdots \frac{d\gamma
_8}{2\pi } \exp[\frac{\beta }{16}A(\gamma )
+ i\sum_{i,j=1}^8\gamma _j (S^{-1})_{ji}m^i], \end{eqnarray}
where we have introduced the function

\begin{equation} A(\gamma ) = \sum _{j=1}^8\cos \gamma _j + \sum _
{\alpha =9}^{128}\epsilon _\alpha \cos [\sum _{j=1}^8\gamma
_j(S^{-1}T)_{j\alpha }]. \end{equation}
In order to fulfill the conditions (3.4) we put

\begin{equation} m^i = 2 \tilde{m}^i + f \mbox{, with $\tilde{m}^i$
integer, and $f=0$ or 1}. \end{equation}
For our choice of the transformation, the $f$-dependence of
(3.8) becomes particularly simple and only involves $\gamma _8$.

We now reintroduce the suppressed indices $p_{\rho \nu }$ into (3.8),
(3.10), and insert the result into the partition function (2.18).
We find

\begin{eqnarray}  Z[J] =
& & \sum_{\hat{n}}W_{\hat{n}} \sum_{\tilde{m}_{\rho \nu }^i(p)}
\sum_{f_{\rho \nu }(p)} \prod_{p_{\rho \nu }} \left( \int
\frac{d\gamma _1}{2\pi }\cdots \frac{d\gamma _8}{2\pi }\right.\\
& & \left.\exp\left[ \frac{\beta }{16}A(\gamma ) + 2i\sum_{i,j=1}^8
\gamma _j (S^{-1})_{ji} \tilde{m}_{\rho \nu }^i(p) + i \gamma
_8f_{\rho \nu }(p)\right] \right) \prod_{r\mu a}\delta ^a[C_\mu
^a(r;\tilde{m},f,J,\hat{s} )].\nonumber \end{eqnarray}
The constraints (2.17) now depend upon $\tilde{m}^i$ and $f$. For the
formulation it is convenient to extend $f_{\mu \nu }(r)$ to an
antisymmetric matrix. Thus $f_{\mu \nu }(r) =0,1$ for $\mu <\nu $, and
$f_{\mu \nu }(r) =0,-1$ for $\mu >\nu $. The constraints then become

\begin{eqnarray} C_\mu ^a(r)
& = & 2\sum_{\nu >\mu }
[\tilde{m}_{\mu \nu }^{(8,1,2)}(r) - \tilde{m}_{\mu \nu
}^{(6,5,4)}(r-\nu )] - 2\sum_{\nu <\mu }[\tilde{m}_{\nu \mu
}^{(8,7,6)}(r) - \tilde{m}_{\nu \mu }^{(2,3,4)}(r-\nu )] \nonumber\\
& + & \sum_{\nu \neq \mu } \Delta_\nu f_{\mu \nu }(r) +
J_\mu (r)\hat{s}_{\mu \hat{n}}^a(r), \end{eqnarray}
where $\Delta _\nu $ denotes the left lattice derivative. The tensor
$f_{\mu \nu }(r)$ will be recognized as the $Z_2$ field strength
tensor. As in (2.17) one has to use the first, second, or third upper
index of $\tilde{m}_{\mu \nu }$ for $a=\psi ,\vartheta ,\varphi $.

There is a symmetry relation in (3.11) which underlines the importance
of the $Z_2$ tensor $f_{\rho \nu }(p)$. Replace $\beta \rightarrow
-\beta $ and substitute $\gamma _j \rightarrow \gamma _j + \pi $ for
all $j$ (the integrand is periodic). Using the definition of $A(\gamma
)$ in (3.9) and the fact that $\sum_{j=1}^8(S^{-1}T)_{j\alpha }=1$ for
all $\alpha $, one finds that $A(\gamma )$ reverses sign, i.e. $\beta
A(\gamma )$ stays invariant. The second term in the exponent of (3.11)
changes by a multiple of $2\pi i$, because $\sum_{j=1}^8(S^{-1})_{ji}
= \delta _{8i}$, and $\tilde{m}_{\rho \nu }^8(p)$ is integer. Finally
the third term changes by $i\pi f_{\rho \nu }(p)$. In this way one
finds that the bracket $(\cdots )$ in (3.11) is even in $\beta $ for
$f_{\rho \nu }(p)=0$, and odd in $\beta $ for $f_{\rho \nu }(p)= 1$.

Up to this point all formulae were exact.
It appears tempting now to proceed as follows in the continuum limit
of large $\beta $. If the function $A(\gamma )$ has an isolated
maximum, the integrals over $\gamma _j$ are dominated by the region
where $A(\gamma )$ becomes maximal, and the integrations over the
$\gamma _j$ can be extended to the interval from $-\infty$ to
$\infty$. A quadratic expansion around the maximum would then lead to
gaussian integrals. If all the $\epsilon _\alpha $ were equal to 1, we
would indeed have a simple maximum at $\gamma _j=0$ for all $\gamma
_j$. This would correspond to the situation in the abelian theory and
to the replacement $I_l(z) \rightarrow e^z e^{-l^2/2z}/\sqrt{2\pi z}$
which was used in \cite{bmk} (these authors used by mistake
$e^{-l^2/4z}$ instead of the correct $e^{-l^2/2z}$, which has,
however, no consequences there).

In our case the different signs of the $\epsilon _\alpha $ change the
situation drastically. One finds that the function $A(\gamma )$
assumes its maximal value of 16 if $\gamma _1,\gamma _2,\gamma _3$ are
arbitrary, and $\gamma _j = 0$ for $j=4,\cdots ,8$. Even a quadratic
approximation in $\gamma _j$ for $j=4,\cdots,8,$ and fixed $\gamma
_1,\gamma _2,\gamma _3$ is not possible because the matrix of the
second derivatives has two zero eigenvalues. For $\gamma _1$=$\gamma
_2$=$\gamma _3$=$0$ even three eigenvalues vanish. So it would be
necessary to go to a higher order in the expansion; but then the
integrations are no longer gaussian and cannot be performend. This is
the way in which non abelian gauge theory protects itself from being
solved analytically!

Nevertheless the present formalism will clearly show, how the area law
for large loops arises. We keep $\beta $ arbitrary, not necessarily
large, and first solve the constraints.

\setcounter{equation}{0}\addtocounter{saveeqn}{1}%

\section{Solution of the constraints}

It is convenient to rewrite the $\vartheta $-constraints in the form

\begin{equation} \delta ^\vartheta [C_\mu ^\vartheta (r)] =
\sum_{k_\mu (r)=-\infty}^\infty \delta ^\vartheta [4k_\mu (r)] \delta
_{C_\mu ^\vartheta (r)-4k_\mu (r),0}. \end{equation}
This introduces additional sums over the $k_\mu (r)$, and factors
$\delta ^\vartheta [4k_\mu (r)]$. The $\vartheta $-constraints now
also appear in form of a Kronecker-$\delta $.

Let us first consider the constraints modulo 2, which obviously only
concerns the second line of (3.12). The factors $\hat{s}_{\mu
\hat{n}}^a(r) = \pm 1$ may be dropped, and for all three cases $a=\psi
,\vartheta ,\varphi $ we find the equations

\begin{equation} \sum_\nu \Delta_\nu f_{\mu \nu }(r) + J_\mu (r)
= 0 \mbox{\quad (mod 2) for all\quad }r,\mu . \end{equation}
These are identical to the equations for $l_{\mu \nu }(r)$ in the
abelian case, except that they are equations modulo 2. They show
already the appearance of a $Z_2$ structure. We concentrate on a
geometrical formulation of the solution. Recall that $f_{\mu \nu
}(r)=0,1$ for $\mu <\nu $ and that $f_{\mu \nu }(r)$ is antisymmetric.
Therefore it is convenient to use the symbol $\epsilon _{\mu \nu } =
(1,-1,0)$ for $(\mu <\nu ,\mu >\nu ,\mu =\nu )$. Let $S$ be a two
dimensional surface, and

\begin{equation} f_{\mu \nu }^{(S)}(r) = \left\{ \begin{array}{l}
\epsilon _{\mu \nu }\mbox{\quad if the plaquette }r_{\mu \nu }\mbox{
is part of the surface $S$,}\\ 0\mbox{\quad otherwise.} \end{array}
\right. \end{equation}
The following statements hold:
\begin{itemize}
\item If $S$ has the Wilson loop as boundary, then $f_{\mu \nu
}^{(S)}(r)$ is a solution of (4.2).
\item If $S$ is a closed surface, then $f_{\mu \nu }^{(S)}(r)$ is a
solution of the homogeneous equation $\sum_\nu \Delta_\nu f_{\mu \nu
}^{(H)}(r) = 0$ (mod 2). The most general solution can be obtained as
a superposition of a special solution $f_{\mu \nu }^{(S)}(r)$ with $S$
bounded by the loop, and a sum over solutions of the homogeneous
equation.
\end{itemize}
The proof is obvious. Equation (4.2) involves exactly all the
plaquettes which contain the link $r_\mu $. Links $r_\mu $ on the loop
appear in an odd number of plaquettes of the associated surface $S$,
while links $r_\mu $ which are not part of the loop appear in an even
number (including 0) of plaquettes of $S$.

In the following we will restrict the discussion to planar loops for
simplicity. A solution of special importance is the layer belonging to
the minimal surface of the Wilson loop $W$,

\begin{equation} f_{\mu \nu }^{(min)}(r) = \left\{
\begin{array}{l} \epsilon _{\mu \nu }\mbox{\quad if the plaquette
}r_{\mu \nu }\mbox{ is part of the minimal surface,}\\ 0\mbox{\quad
otherwise.} \end{array} \right. \end{equation}
The $f_{\mu \nu }^{(min)}(r)$ associated with the minimal surface
fulfills (4.2) exactly, not only modulo 2, if the loop is oriented
appropriately. For different surfaces, on the other hand, this is not
true.

The general solution of the homogeneous equation $\sum_\nu \Delta_\nu
f_{\mu \nu }^{(H)}(r) = 0$ (mod 2) can be written down explicitly. It
depends upon the dimension $d$. In order to guarantee the antisymmetry
of $f_{\mu \nu }(r)$ we introduce the symbol (mod$_{\mu \nu }$ 2); it
is identical with (mod 2) for $\mu <\nu $, but reverses sign for $\mu
>\nu $. One then has

\begin{eqnarray} f_{\mu \nu }^{(H)}(r) = \left\{\begin{array}{r}
\sum_\lambda \epsilon _{\mu \nu \lambda }\Delta _\lambda f(r)
\mbox{\quad (mod$_{\mu \nu }$ 2) for $d=3$,}\\
\sum_{\lambda \kappa }\epsilon _{\mu \nu \lambda \kappa }\Delta
_\lambda f_\kappa (r) \mbox{\quad (mod$_{\mu \nu }$ 2) for
$d=4$.}\end{array} \right. \end{eqnarray}
The function $f(r)$ in three dimensions is unique up to a constant.
For $d=4$ one has a gauge freedom, i.e. adding a gradient
$\Delta_\kappa \Lambda (r)$ to $f_\kappa (r)$ will not change
$f_{\mu \nu }^{(H)}(r)$. The simplest way to remove this ambiguity
is to choose an axial gauge by imposing $\sum_\kappa n_\kappa f_\kappa
(r) =0$ (mod 2). In both cases the values of $f(r)$ and $f_\kappa
(r)$, respectively, are restricted to 0 and 1.

Switching from one surface $S$ to another $S'$ for the special
solution can also be rephrased in terms of the solution of the
homogeneous equation. In $d=3$ dimensions it corresponds to changing
$f(r)$ by 1 inside the volume between the two surfaces. (The use of
the left derivative in (4.2), (4.5) specifies which points of the
surface have to be considered as inside or outside). For $d=4$ one has
to choose a three-dimensional volume spanned by the surfaces with,
roughly speaking, normal vector in $\kappa $-direction at the point
$r$. One then has to change $f_\kappa (r)$ by 1 inside the volume, and
subsequently transform to the axial gauge.

The essential part of the constraints has now been solved. We put

\begin{equation} f_{\mu \nu }(r) = f_{\mu \nu}^ {(min)}(r) +
\sum_{\lambda \kappa }\epsilon _{\mu \nu \lambda \kappa } \Delta
_\lambda f_\kappa (r) \mbox{\quad (mod$_{\mu \nu }$
2).}\end{equation}
The minimal surface layer $f_{\mu \nu}^ {(min)}(r)$, defined in
(4.4), is no longer a variable, but uniquely fixed by the loop.
The $f_\kappa (r)=0,1$ are unconstrained. The index
$\kappa $ on $\epsilon _{\mu \nu \lambda \kappa },f_\kappa (r)$ and in
the sum appears for $d=4$ only. For $d=3$ it has to be dropped, here
and wherever it appears in subsequent formulae.

We next introduce the solution (4.6) into (3.12). The second line
is now definitely even, therefore we denote it by

\begin{equation} 2R_{\mu \hat{n}}^a(r) \equiv \sum_\nu \Delta_\nu
f_{\mu \nu }(r) + J_\mu (r)\hat{s}_{\mu \hat{n}}^a(r). \end{equation}
For later use we specify the variables upon which $R_{\mu
\hat{n}}^a(r)$ can depend. The term $J_\mu (r)\hat{s}_{\mu
\hat{n}}^a(r)$ is strictly local, i.e. only depends on the argument
$r$. The $f_\kappa (r)$, on the other hand, appear as $\Delta _\nu
\Delta _\lambda f_\kappa (r)$ with $\nu \ne \lambda $. Therefore they
enter also with shifted arguments $r'$. The points $r'$ and $r$ are
neighbors in the sense that all components of $r-r'$ are either 0 or
1. Finally, one has to note that (4.6) is only an equation modulo 2.
Therefore $\sum_\nu \Delta _\nu f_{\mu \nu }(r)$ as well as $R_{\mu
\hat{n}}^a(r)$ can also depend on the minimal layer $f_{\mu \nu
}^{(min)}(r)$.

The constraints (3.12) now become

\begin{eqnarray} C_\mu ^a(r) =
& &  2\sum_{\nu >\mu } [\tilde{m}_{\mu \nu
}^{(8,1,2)}(r) - \tilde{m}_{\mu \nu }^{(6,5,4)}(r-\nu )] - 2\sum_{\nu
<\mu }[\tilde{m}_{\nu \mu }^{(8,7,6)}(r) - \tilde{m}_{\nu \mu
}^{(2,3,4)}(r-\nu )] \nonumber\\
& & + 2R_{\mu \hat{n}}^a(r) \stackrel{!}{=}4\delta ^{a\vartheta }
k_\mu (r). \end{eqnarray}
For $a=\psi $ (first upper index $i$ on $\tilde{m}^i$) and
$a=\varphi $ (third upper index $i$ on $\tilde{m}^i$), i.e. for the
even indices $i$, the index $i$
appears in both sums of (4.8). For a fixed $r$ one has $2d$ linear
equations (corresponding to $a=\psi ,\varphi $ and $\mu =1,\cdots ,
d$) for $4 d(d-1)/2$ quantities $\tilde{m}_{\mu \nu }^{(2,4,6,8)}$.
These equations are not independent due to the identity

\begin{equation} \sum_\mu [C_\mu ^\psi (r)- C_\mu ^\varphi (r-\mu )] =
2 \sum_\mu [R_{\mu \hat{n}}^\psi (r) - R_{\mu \hat{n}}^\varphi (r-\mu
)] . \end{equation}
The rhs of (4.9) vanishes because $f_{\mu \nu }$ is antisymmetric, and
because the neighboring projectors in the loop have to coincide as
mentioned in sect. 2. We checked explicitly for $d=3,4$ that the
equations may be simply used to eliminate some of the $\tilde{m}_{\mu
\nu }^i(r)$. Any eliminated $\tilde{m}_{\mu \nu }^i(r)$ depends
linearly on other unconstrained $\tilde{m}_{\mu '\nu '}^{i'}(r')$
and on $R_{\mu '\hat{n}}^a(r')$, where the components of $r'$-$r$ are
either 0 or $\pm 1$.

For $a=\vartheta $ (second upper index $i$ on $\tilde{m}^i$), i.e. for
the odd indices $i$, the situation is even simpler. Each index $i$
enters only in one of the sums in (4.8), it is convenient to eliminate
some of the $\tilde{m}_{\mu \nu }^1(r)$ and $\tilde{m}_{\nu \mu
}^7(r)$.

Finally we can write the solutions of the constraints in the following
form, which eliminates some of the $\tilde{m}_{\mu \nu }^i(r)$,
leaving the rest unconstrained.

\begin{equation} \tilde{m}_{\mu \nu }^i(r) = L_{\mu \nu }^i
[\tilde{m}_{\mu '\nu '}^{i'}(r'),R_{\mu '\hat{n}}^a(r')] + 2k_{\mu \nu
}^i(r), \end{equation}
with $k_{\mu \nu }^i(r) = (k_\mu(r),k_\nu(r),0)$ for
$(i=1,i=7,$otherwise).
The $L_{\mu \nu }^i$ are linear combinations of their arguments,
only coefficients $0,\pm 1$ appear. The arguments $r,r'$ are neighbors
in the sense explained before. The loop current $J_\mu (r)$ enters
only in $R_{\mu '\hat{n}}^a(r')$.

Note the drastic difference in the type of the constraints (4.2) (or
the corresponding constraints in eq. (6) of ref \cite{bmk} for the
$U(1)$ case) on one hand, and the constraints (4.8) just considered on
the other. The former involve a difference operator applied to one
plaquette variable $f_{\mu \nu }(r)$, the corresponding Green function
being non-local and coupling the solution to the current over a long
range. In contrast, the latter constraints involve several plaquette
variables $\tilde{m}_{\mu \nu }^i(r)$ and can just be used to
eliminate some of these. This elimination leads to an almost local
coupling to the current, involving neighbors only.

\setcounter{equation}{0}\addtocounter{saveeqn}{1}%

\section{Confinement}

The essential feature, which finally arose in our formulation, is the
presence of the $Z_2$ field strength tensor $f_{\mu \nu }(r)$ which
obeys the field equation (4.2). The solutions of this equation can be
characterized by two-dimensional surfaces; a layer with the Wilson
loop as boundary, possibly together with closed surfaces. One may
expect that the presence of such a layer will lead to an area law.
For a qualitative understanding of the confinement mechanism we use
the explicit form (4.6) for the solution of the field equation (4.2).
It contains the fixed layer $f_{\mu \nu }^{(min)}(r)$, together
with the unconstrained $Z_2$ variable $f_\kappa (r)$. The solutions of
the remaining constraint equations (4.8) for the $\tilde{m}_{\mu \nu
}^i(r)$ have the form (4.10).

Consider now the expression (3.11) for the partition function $Z[J]$,
and introduce the solutions (4.6), (4.10) of the constraints into the
exponential on the rhs. This gives

\begin{eqnarray} \lefteqn{2\sum_{i,j=1}^8\gamma
_j(S^{-1})_{ji}\tilde{m}_{\rho \nu }^i(p) + \gamma _8f_{\rho \nu
}(p) }\nonumber\\
& = & 2\sum_{i,j=1}^8\gamma _j(S^{-1})_{ji}\{L_{\rho \nu
}^i[\tilde{m}_{\rho '\nu '}^{i'}(p'),R_{\rho
'\hat{n}}^a(p')] + 2k_{\rho \nu }^i(p)\}\\
& + & \gamma _8\{f_{\rho \nu }^{(min)}(p) +
\sum \epsilon _{\rho \nu \lambda \kappa }\Delta _\lambda f_\kappa
(p) \mbox{\quad (mod$_{\mu \nu }$ 2)}\}.\nonumber \end{eqnarray}
The Wilson loop enters into this expression in two different ways.
First there is a dependence on the current $J_{\rho '} (p')
\hat{s}_{\rho '\hat{n}}^a(p')$ which arises from the second term of
$R_{\rho '\hat{n}}^a(p')$ in (4.7). Secondly there is a dependence on
the minimal layer $f_{\rho \nu }^{(min)}(p)$ which enters into the
first term of $R_{\rho '\hat{n}}^a(p')$, as well as explicitely in the
factor of $\gamma _8$.

The current $J$ is present on a one-dimensional set, the
minimal layer $f^{(min)}$ on a two-dimensional set.
Besides this, both quantities enter in a quite similar way into (5.1).
One may therefore expect, that for large loops the
dependence on the one-dimensional current $J$ can be neglected
compared to the dependence on the two-dimensional layer $f^{(min)}$.

If we neglect the dependence on $J_{\rho '} (p')$ the partition
function becomes independent of the $\hat{s}_{\rho '\hat{n}}^a(p')$
and the sum $\sum_{\hat{n}}W_{\hat{n}} = 2$ can be performed. Assuming
that the $\gamma $-integrations have been done, the degrees of freedom
are now in the remaining unconstrained $\tilde{m}_{\rho \nu }^i(p) =
-\infty,\cdots,\infty$, the $f_\kappa (p)=0,1$, and the $k_\mu (r)
=-\infty,\cdots,\infty$ introduced at the beginning of sect. 4.

The discussion of (4.8) showed that the solutions couple neighbors
only. This means that (5.1), which appears in the exponential in
(3.11), only depends on these variables with arguments
$p,p',p''$; here $p',p''$ are neighbors in the sense that all
components of $p'$-$p$ are $0,\pm 1$, all components of $p''$-$p$ are
$0,\pm 1,\pm 2$.

We digress for a technical point. Neither the exponential
in (3.11) with it's complex argument, nor the factors $\delta
^\vartheta [4k]$ are positive definite. Actually, according to the
definition (2.16), one has $\sum_k\delta ^\vartheta [4k]=0$, because
the Haar measure fulfills $H(0) = 0$. If desired, one could bring the
expression into the usual form of a partition function with positive
summands, by performing a twofold partial summation with respect to
the $k_\mu (r)$.

The whole loop dependence is now in the $f_{\rho \nu }^{(min)}(p)
$ belonging to the minimal surface. It acts like a space-time
dependent external field, comparable, say, to a constant magnetic
field switched on in a finite volume of an Ising model. $Z[J]$ is a
partition function where the variables couple to neighbors only,
a well known standard situation in statistical mechanics. For large
subsystems it therefore factorizes into products refering to the
subsystems and, correspondingly, has an exponential dependence on the
volume. This is, of course, nothing else but the fact that the
free energy is an extensive quantity. Rigorous proofs, which apply for
any dimension, can be found in \cite{fac}.

Consider now a loop $0< x_1\le R,0<x_4\le T$ in the $x_1$-$x_4$-plane
for definiteness, with $R$ and $T$ large. We divide the
$x_1$-$x_4$-plane inside, as well as outside of the loop, into
rectangles; these rectangles are then extended to $d$-dimensional
boxes into the orthogonal directions. This means that we define
regions $V^{(n)}$ by the inequalities $r_1^{(n)} <x_1\le
r_2^{(n)},t_1^{(n)} <x_4\le t_2^{(n)},x_2,x_3$ arbitrary. The
rectangle $r_1^{(n)} <x_1\le r_2^{(n)},t_1^{(n)} <x_4\le t_2^{(n)}$
has to lie either completely inside the loop, or completely outside
the loop. If not only $R$ and $T$, but also all the differences
$r_2^{(n)} - r_1^{(n)}$ and $t_2^{(n)} - t_1^{(n)}$ are large, the
partition function will factorize,

\begin{equation} Z[J] = \prod_n Z^{(n)}.  \end{equation}
Consider now the ratio $Z[J]/Z[0]$, with $Z[0]$ the expression without
loop. Obviously all the outer factors cancel. For the inner ones, on
the other hand, one has $f_{\rho \nu }^{(min)}(p) =1$ in
the numerators, but 0 in the denominators. Thus the ratios
are different from 1. Because of the factorization property, the
volumes of the regions $V^{(n)}$ have to enter in the exponent. This
finally implies that the area $A = R\times T$ of the loop enters in
the exponent, so the result may be written as

\begin{equation} Z[J]/Z[0] = \exp[-\sigma A].\end{equation}
We have obtained the area law for large loops.

Several comments are appropriate here.

First one may wonder what would happen with our argumentation, if we
would replace $f_{\rho \nu }^{(min)} (p)$, associated with the minimal
surface, by a solution $f^{(S)}_{\rho \nu }(p)$, belonging to a
different surface $S$. Obviously the simplicity of the situation for
the regions $V^{(n)}$ inside and outside would break down, and
factorization would not lead to a simple relation. The minimal surface
is really unique for the argumentation.

The neglection of the dependence on $J$ would certainly have been
wrong if performed in the original expression described by the Euler
angles $\psi _\mu (r),\vartheta _\mu (r),$ $\varphi _\mu (r)$. There
one had only the current $J$ but no layer $f^{(min)}$ showed up.
Therefore confinement has to arise from the dependence on $J$ in a
complicated way. In our formulation the formalism led to another
quantity, the minimal layer $f^{(min)}$. This appears as the natural
quantity which describes the long distance physics and dominates the
residual direct dependence on $J$.

For illustration one can have a look on the strong coupling limit.
According to the discussion at the end of sect. 3, the bracket
$(\cdots )$ in (3.11) is even in $\beta $ for $f_{\rho \nu }(p)=0$,
and odd in $\beta $ for $f_{\rho \nu }(p)=\epsilon _{\rho \nu }$.
Therefore the order $\beta ^0$ only contributes outside the surface
layer, while the order $\beta $ terms come from plaquettes on the
surface layer. In this way we recover the well known lowest order
strong coupling result $Z[J]/Z[0]\sim \beta ^A$. More important, we
have seen that indeed $f^{(min)}$ is the crucial quantity.

For a Wilson loop in the adjoint representation one does not expect an
area law, because the charges can be screened by pair creation. This
can be easily checked in our approach. The traces in the adjoint and
in the fundamental representation are related by
$TrW_{(1)}=[TrW_{(1/2)}]^2-1$. With our parametrization we obtain

\begin{eqnarray} [TrW_{(1/2)}]^2 & = & \left[ \sum_{\hat{n}}
W_{\hat{n}} \exp [i\sum_{q\lambda b}J_\lambda (q)\hat{s}_{\lambda
\hat{n}}^b(q)\Theta _\lambda ^b(q)]\right] ^2 \nonumber\\
& = & \sum_{\hat{n}\hat{n}'} W_{\hat{n}}W_{\hat{n}'} \exp \left[
i\sum_{q\lambda b}J_\lambda (q)[\hat{s}_{\lambda \hat{n}}^b(q)
+\hat{s}_{\lambda \hat{n}'}^b(q)] \Theta _\lambda ^b(q)\right].
\end{eqnarray}
The last term on the rhs of the modified equation (3.12) becomes
$J_\mu (r)[\hat{s}_{\mu \hat{n}}^a(q)+\hat{s}_{\mu \hat{n}'}^a(q)]$
and is always even. Therefore (4.2) becomes a homogeneous equation, no
minimal layer and no area law will appear. Similarly one can see that
we don't get confinement if we replace $SU(2)$ by $SO(3)$.

With some technical effort or a more streamlined approach it should be
possible to carry through a similar analysis for $SU(3)$. It would be
nice to see, how the formalism would create the expected $Z_3$
structure.

Our conclusions which led to the area law would break down if the
result, by some reason whatsoever, would be independent of
$f_{\rho \nu }(p)$, thereby giving a vanishing string tension. This
appears hardly possible for a finite lattice. We have seen before that
there is indeed an essential dependence on $f_{\rho \nu }(p)$ in the
strong coupling limit $\beta \rightarrow 0$. Such a dependence must
survive for all finite $\beta $ because the original expression $Z[J]$
in (2.1) clearly fulfills the strict inequalities $0<Z[J]<Z[0]$. The
string tension might, however, vanish in a certain region of $\beta $
after performing the thermodynamic limit. In particular such an effect
could be expected in higher dimensions, where the presence of the two
dimensional layer becomes relatively less important than in lower
dimensions. Indeed it is known \cite{fuenf} that lattice $SU(2)$ has a
first order phase transition for $d=5$ at $\beta _c=1.642 \pm 0.015$.

We come back to $d=4$. At the end one is interested in the continuum
limit $\beta \rightarrow \infty $ which requires a particular
investigation. If the string tension is a physical quantity
and $\beta $ goes to infinity as prescribed by the renormalization
group, a non vanishing string tension for the lattice theory will
persist in the continuum limit.

\setcounter{equation}{0}\addtocounter{saveeqn}{1}%

\section{Interpretation and conclusions}

There is an extensive literature on the various pictures of
confinement which cannot be discussed here. For a recent review we
refer to \cite{sim}. We come directly to the physical interpretation
of our results. The key is equation (4.2) for the $Z_2$ field strength
tensor,

\begin{equation} \sum _\nu \Delta _\nu f_{\mu \nu }(r) + J_\mu
(r) = 0 \mbox{\quad (mod 2)}. \end{equation}
The solutions in form of layers on two-dimensional surfaces were
discussed in detail in sect. 4.

In $d$=3 dimensions put $f_{\mu \nu }(r) = \sum _\lambda \epsilon
_{\mu \nu \lambda }B_\lambda (r) \mbox{\quad (mod$_{\mu \nu }$ 2)}$.
Then (6.1) becomes $\nabla \times {\bf B} = {\bf J}$ (mod 2). The
magnetic field ${\bf B}$ has sources corresponding to magnetic
monopoles. It is reasonable to use the right derivative in the
divergence, and to associate the monopole density $\tilde{\rho }$ with
cubes as usual. We therefore define

\begin{equation} \tilde{\rho }(r_{123}) = \sum_\nu \Delta _\nu
^{(right)} B_\nu (r) \mbox{\quad (mod 2)}.  \end{equation}
The solution $f_{\mu \nu }^{(min)}(r)$ then immediately leads to a
double layer of monopoles in the cubes on both sides of the
minimal surface.

For $d$=4 we define the dual tensor $\tilde{f} _{\mu \nu }(r) =
(1/2)\sum _{\lambda \kappa } \epsilon _{\mu \nu \lambda \kappa }f
_{\lambda \kappa }(r) \mbox{ (mod$_{\mu \nu }$ 2)}$. The conserved
(mod 2) magnetic current $\tilde{J}_\mu$ lives on 3-dimensional cubes
$r_{\rho \lambda \kappa }$, where $\rho ,\lambda ,\kappa $ denote the
three directions orthogonal to $\mu $.

\begin{equation} \tilde{J}_\mu (r_{\rho \lambda \kappa }) = \sum _\nu
\Delta_\nu ^{(right)}\tilde{f}_{\mu \nu }(r) \mbox{\quad (mod 2)}.
\end{equation}
Consider a loop in the $x_1$-$x_4$-plane, with $x_4$ interpreted as
euclidean time. Let $x_1,x_4$ be within the loop and suppress the
$x_4$-extension of the cubes. For the solution $f ^{(min)}_{\mu \nu
}(r)$ we then get a non-vanishing $\tilde{J}_\mu $ on all
plaquettes in the $x_1$-$x_2$-plane and in the $x_1$-$x_3$-plane which
contact the line $x_2$=$x_3$=0. We thus have a
string of electric field $E_1(r) = f_{14}(r)$ in
$x_1$-direction, concentrated on $x_2$=$x_3$=0. This is surrounded by
magnetic current loops parallel to the $x_2$-$x_3$-plane which circle
around the electric string. The configuration is therefore just dual
to an Abrikosov vortex in a type II superconductor, where the magnetic
field is encircled by the electric current. Flux quantization is
evident, the $Z_2$-structure only allows for one unit of flux.

Configurations $f_{\mu \nu }(r)$ in (4.6) with $f_\kappa
(r)\ne 0$ belong to other surfaces which are bounded by the loop as
discussed in sect. 4. In addition closed surfaces can appear. The
interpretation is similar as above. For illustration, connect e.g.
two points in the $x_1$-$x_2$-plane by a path in form of a stair. Then
$\tilde{J}_\mu $ lies on the plaquettes which point from the stair
into positive and negative $x_3$-direction. All surfaces are summed
with the appropriate weight in the partition function. A careful
investigation of the various weights should give information about the
extension of the electric flux tube.

We finally check the dual London equation, $\nabla \times \tilde{{\bf
J}} =\frac{1}{\Lambda }{\bf E}$\quad (mod 2). It is easily seen that
$(\nabla \times \tilde{{\bf J}})_\mu (r)$\quad (mod 2) is equal to 0
(1) if the link $r_\mu $ has contact to an even (odd) number of
plaquettes with non-vanishing magnetic current. For the examples
discussed above this means that $\nabla \times \tilde{{\bf J}}$ runs
along the boundary of the set of plaquettes which carry the magnetic
current. For the minimal layer, $\nabla \times \tilde{{\bf J}}$
is parallel to ${\bf E}$ as it should. It is, however, not
concentrated on $x_2$=$x_3$=0 as the electric field, but on the four
lines $x_2$=0, $x_3=\pm 1$ and $x_3$=0, $x_2=\pm 1$. For the stair,
$\nabla \times \tilde{{\bf J}}$ is shifted by $x_3=\pm 1$ with respect
to ${\bf E}$. In general the dual London equation is essentially
fulfilled, the two sides of the equation are just slightly shifted
against each other. This might be interpreted by a non-vanishing
Ginzburg-Landau coherence length $\tilde{\xi }$ which leads to a
``normal'' region near the string, where the London equation is not
valid.

Let us compare with some familiar types of monopoles in the
literature.

The charges of the $U(1)$ monopoles in \cite{bmk} can take all
integers, in obvious contrast to our $Z_2$ structure which only allows
0 and 1.

A popular definition of monopoles in $SU(2)$ is discussed e.g. in
\cite{tom}. Let $\eta _{(p)} \equiv sign\:Tr\:U_{(p)}$ denote the sign
of the plaquette action, and $\eta _c=\prod_{p \in \partial c}\eta
_{(p)}$ the product of the $\eta _{(p)}$ around the boundary of the
cube $c$. Then $\eta _c=-1$ represents a monopole in the (space like)
cube $c$. There is a $Z_2$ structure as in our case.

Another frequently applied definition, reviewed e.g. in \cite{hay},
uses the maximal abelian gauge. In a first step one maximizes the
quantity $R=\sum_{r\mu }Tr[\sigma _3U_\mu (r)\sigma _3U_\mu ^+(r)]$.
The link matrices are then decomposed into a non-abelian and a $U(1)$
part, e.g. one can take the abelian link angle as the
phase of $[U_\mu (r)]_{11}$. The $U(1)$ monopoles are then defined
according to the DeGrand Toussaint construction \cite{dgt} which
allows monopole charges $0,\pm 1,\pm 2$.

The monopoles which naturally arose in the present work have no
direct relation to any of these. An unconventional
feature of our approach is the presence of monopoles in every
configuration. In the approaches mentioned above there are plenty of
configurations without any monopoles, namely those near the
perturbative vacuum. In our case there is always a surface $S$ bounded
by the loop. This is associated with an electric string and
accompanied by monopole vortices. From a physical point of view this
appears quite attractive.

\quad

\noindent
{\bf Acknowledgement:} I thank I. Bender and H. J. Rothe for reading
the manuscript and for valuable criticism.

\renewcommand{\theequation}{\Alph{section}.\arabic{equation}}
\begin{appendix}
\setcounter{equation}{0}\addtocounter{saveeqn}{1}%
\section{Appendix}

For definiteness we give here the matrix $S$ used by us in sect. 3
when selecting a convenient subset of the ${\bf S}_n$. It reads

\begin{equation} S = \left|\begin{array}{rrrrrrrr}
-1&-1&-1&1&-1&1&1&1\\
-1&-1&1&-1&1&-1&1&1\\
-1&1&1&1&-1&1&1&1\\
-1&-1&1&1&1&1&1&1\\
-1&1&1&-1&-1&1&-1&1\\
-1&-1&-1&1&-1&1&-1&1\\
-1&-1&1&-1&-1&1&-1&1\\
1&1&1&1&1&1&1&1\\
\end{array}\right|  \end{equation}
Recall that the columns of $S$ consist of 8 of the vectors ${\bf
S}_n$, with the property that $\epsilon _n=+1$. The last component,
corresponding to $s_8$, was fixed to 1.
\end{appendix}


\begin{thebibliography}{99}

\bibitem{bmk} T. Banks, R. Myerson, J. Kogut, Nucl. Phys. B 129, 493
(1977).

\bibitem{vil} J. Villain, Journal de Physique, 36, 581 (1975).

\bibitem{bj} I. R. C. Buckley, H. F. Jones, Phys. Rev. D 45, 654
(1992).

\bibitem{bg} I. Bender, D. Gromes, Z. f. Physik C 73, 721 (1997),
hep-lat/9604022.

\bibitem{fac} D. Ruelle, Statistical Mechanics: Rigorous Results, World Scientific, Singapore 1999;
R. B. Griffiths, in Phase Transitions and Critical Phenomena, Vol 1,
ed. C. Domb, M. S. Green, Academic Press, London, New York 1972;
D. Ruelle, Thermodynamic Formalism, Encyclopedia of Mathematics and
it's applications, Vol. 5, ed. G. Gallavotti, Addison-Wesley Publ.
Comp., Reading Mass. 1978.

\bibitem{fuenf} M. Creutz, Phys. Rev. Lett. 43, 553 (1979); M. Baig,
A. Cuervo, Nucl. Phys. B (Proc. Suppl.) 4, 21 (1988).

\bibitem{sim} Yu. A. Simonov, Physics-Uspekhi 39(4), 313 (1996),
hep-ph/9709344.

\bibitem{tom} E. T. Tomboulis, Phys. Lett. B 303, 103 (1993).

\bibitem{hay} R. W. Haymaker, Proc. of the Int.
School of Physics ``Enrico Fermi'', Course 80, Selected topics in non
perturbative QCD, Varenna, 27 June - 7 July 1995, ed. A. Di Giacomo,
D. Diakonov, IOS press 1996, hep-lat/9510035.

\bibitem{dgt} T. A. DeGrand, D. Toussaint, Phys. Rev. D 22, 2478
(1980).

\end{thebibliography}
\end{document}